\documentstyle[12pt]{article}

\begin{document}
\newcommand{\YM}{Yang--Mills\ }
\newcommand{\N}{$N_{CS}\;$}
\newcommand{\Nz}{$N_{CS}=0\;$}
\newcommand{\No}{$N_{CS}=1\;$}
\newcommand{\ur}[1]{~(\ref{#1})}
\newcommand{\urs}[2]{~(\ref{#1},~\ref{#2})}
\newcommand{\urss}[3]{~(\ref{#1},~\ref{#2},~\ref{#3})}
\newcommand{\eq}[1]{eq.~(\ref{#1})}
\newcommand{\eqs}[2]{eqs.(\ref{#1},\ref{#2})}
\newcommand{\eqss}[3]{eqs.(\ref{#1},\ref{#2},\ref{#3})}
\newcommand{\eqsss}[2]{eqs.(\ref{#1}--\ref{#2})}
\newcommand{\Eq}[1]{Eq.~(\ref{#1})}
\newcommand{\Eqs}[2]{Eqs.(\ref{#1},\ref{#2})}
\newcommand{\Eqsss}[2]{Eqs.(\ref{#1}--\ref{#2})}
\newcommand{\fig}[1]{Fig.~\ref{#1}}
\newcommand{\figs}[2]{Figs.\ref{#1},\ref{#2}}
\newcommand{\figss}[3]{Figs.\ref{#1},\ref{#2},\ref{#3}}
\newcommand{\beq}{\begin{equation}}
\newcommand{\eeq}{\end{equation}}
\newcommand{\e}{\varepsilon}
\newcommand{\ee}{\epsilon}
\newcommand{\la}[1]{\label{#1}}
\newcommand{\ui}{$U_{int}\,$}
\newcommand{\n}{\bf{n}}
\newcommand{\r}{\bf{r}}
\newcommand{\x}{\bf{x}}
\newcommand{\doublet}[3]{\: \left(\begin{array}{c} #1 \\#2
\end{array} \right)_{#3}}
\newcommand{\tha}[3]{\eta_{#1 #2}^{#3}}
\newcommand{\thb}[3]{\bar{\eta}_{#1 #2}^{#3}}

\vspace{2.5cm}

\begin{center} {\Large\bf Non-Abelian Stokes Theorem and Quark-Monopole
Interaction}
\footnote{The extended version is published in:  {\it Nonperturbative
approaches to QCD}, Proceedings of the Internat.  workshop at ECT*, Trento,
July 10-29, 1995, D.Diakonov (ed.), PNPI (1995)} \end{center} \vspace{1cm}

\begin{center} {\large\bf Dmitri Diakonov and Victor Petrov} \\
{\small\it Petersburg Nuclear Physics Institute, Gatchina, St.Petersburg
188350, Russia}
\end{center}

\vspace{.5cm}

\thispagestyle{empty}

\abstract{We derive a new non-abelian Stokes theorem by rewriting the
Wilson loop as a gauge-invari\-ant area integral, at the price of
integrating over an auxiliary field from the coset $SU(N) / \-
[(U(1)]^{N-1}$ space. We then introduce the relativistic quark-monopole
interaction as a Wess--Zumino-type action, and extend it to the
non-abelian case. We show that condensation of monopoles and
confinement can be investigated in terms of the behaviour of the monopole
world lines. One can thus avoid hard problems of how to introduce
monopole fields and dual Yang--Mills potentials.}

\vspace{1cm}

\section{Introduction}

On the way to describe confinement as due to the monopole condensation
several serious technical problems arise. Until now monopole condensation
has been fully theoretically understood only in theories undergoing
spontaneous breaking of colour symmetry down to the $U(1)$ subgroup(s). We
mean a) the Georgi--Glashow model in 2+1 dimensions by Polyakov
\cite{Pol1} and b) the supersymmetric model in 3+1 dimensions by Seiberg
and Witten \cite{SW}. In both cases there is an elementary Higgs field in
the adjoint representation, whose non-zero vacuum expectation value
mercilessly breakes the gauge group down to the $U(1)$. The compactness
of the remaining $U(1)$ group allows Polyakov--'t Hooft monopoles. In
2+1 dimensions they are `pseudoparticles' and cannot condense; in 3+1
dimensions they can.  In both cases a linear confining potential is
obtained for particles which are electrically charged in respect to the
unbroken $U(1)$ subgroup.

In principle, in QCD a similar scenario could take place: the role of the
elementary Higgs field breaking the gauge symmetry could be played by
some composite gluon operator belonging to the adjoint representation;
its v.e.v. might break the colour $SU(3)$ down to $U(1)\times U(1)$.
However, such a possibility seems to be ruled out by what we know
experimentally: we would have far more hadrons than in reality.
Therefore, one should probably assume that the colour group
remains unbroken -- in contrast to the supersymmetric example of Seiberg
and Witten \footnote{There is an endless philological discussion whether
the Higgs mechanism actually breakes gauge symmetry. Fortunately, there
exists a gauge-invariant formulation, for example: Consider a two-point
correlation function $C(x-y)=\langle \phi^a(x)P_{ab}[x,y]\phi^b(y)\rangle$,
where $\phi^a$ is a scalar (possibly composite) field and $P_{ab}$ is a
P-exponent in the adjoint representation. $C(x-y)$ is perfectly gauge
invariant. If $C(x-y)$ decays exponentially at large separations, there is
no symmetry breaking; if it tends to a constant, the situation is referred
to as Higgs mechanism and spontaneous gauge symmetry breaking; if it decays
as a power of the separation, it can witness the
Berezinsky--Kosterlitz--Thouless phase.}.

An introduction of monopoles without elementary or dynamical Higgs
mechanism is a difficult task by itself. To override the difficulty,
't Hooft~\cite{tH1} has suggested to visualize the $U(1)$ monopoles by
partially fixing the gauge, up to the $U(1)$ transformations. This
procedure is called abelian projection. Monopoles are then objects
similar to those which one would find in the symmetry-broken case.
Of course, such objects are to a great extent dependent on the concrete
choice of the gauge, or the concrete choice of the abelian projection
used. Their desirable condensation is even more obscure: even if it
happens in one gauge, it need not necessarily happen in another
\cite{CPV}. Therefore, it would be helpful to introduce a gauge-invariant
description of monopoles and a gauge-invariant formulation of the
monopole condensation, if there is any. This paper is a step in that
direction.

It should be noted that, if the colour group remains unbroken, the mere
notion of ``monopole condensation" becomes somewhat ambiguous. It seems, at
least, that it {\em can~not} be the usual Higgs mechanism (applied to dual
Yang--Mills potentials), for the same reasons: there would be a
proliferation of hadron states. Indeed, two (dual) gluons out of eight
would be `electrically neutral' in respect to the two $U(1)$ subgroups
where the monopoles live, as well as two new kinds of mesons with the
colour structure $\bar{\psi}\lambda^{3,8}\psi$ and five new kinds of
baryons which happen to have the colour $T_3$ and $Y$ equal to zero. All
these states (multiplied by the variety of quantum numbers) are not
affected by the dual Meissner effect, therefore they are not confined and
should be thus observable as new types of hadrons~\cite{Var}. It is
difficult to imagine that this new realm of hadrons has somehow avoided
registration, the more so that, as recently observed in ref.~\cite{TS}, an
additional discrete symmetry makes some of these unusual hadrons stable
under strong interactions.

If confinement is not due to the dual Higgs effect, then what is due to?
There is increasing evidence from lattice investigations that monopoles
extracted by the maximal abelian projection are important in getting the
area law for the Wilson loop. Therefore, a direct theoretical study of the
relation between monopoles and confinement is highly desirable, without
referring to such evasive quantities as the monopole field and the dual
Yang--Mills potentials. For that reason we prefer the first-quantization
formalism for monopoles  (i.e. path integrals over monopole loops) rather
than the field-theoretical one. Trying to avoid the notion of the dual
fields, we derive the direct (but of course non-local) interaction
between non-abelian charges and non-abelian monopoles.

We start by rewriting the Wilson loop as the ordinary exponent of
a certain flux through the surface spanned on the closed contour of a
heavy quark loop. The price one has to pay for that is an additional
integration over an auxiliary scalar field $\n$ from the
$SU(N)/[U(1)]^{(N-1)}$ coset space. (In the $N=2$ case $\n$ is a
unit 3-vector, ${\bf n}^2 = 1$). We call it the non-abelian Stokes
theorem.  It is a smart formula, in fact. In order to fix the
representation to which the probe quark of the Wilson loop belongs, we
have to add a Wess--Zumino-type term for the $\n$ field. With that
term, the flux becomes that of a gauge-invariant field strength
introduced earlier by Polyakov~\cite{Pol2} and 't Hooft~\cite{tH2} in
relation to monopoles, and the Stokes theorem applies to that particular
field strength. The auxiliary $\n$ field plays the role of the direction
of the elementary Higgs field in colour space.

We next present a relativistic formula for the interaction of a
point charge and a point monopole. It is also a Wess--Zumino-type
formula, but in three dimensions. Using the results of our previous
work \cite{DPWZ} we are able to formulate the quark-monopole interaction
for the non-abelian case.

This paper deals with "kinematical" problems, leaving aside the hard
dynamical one: what is the actual driving force for confinement.
However, we hope that this paper gives a framework for a gauge-invariant
description of the confinement mechanism based on monopole condensation,
whatever it means.

\newpage

\section{Non-abelian Stokes theorem}

The path ordering in the Wilson loop,

\beq
L = \mbox{Tr}\;\mbox{P}\;\exp\left(i \oint A_\mu^a T^a dx_\mu\right),
\la{Wdef}\eeq
can be eliminated at the price of introducing integration over all
gauge transformations along the loop \cite{DP0}. Let $\tau$
parametrize the loop and $A(\tau)$ be the tangent component of the
Yang--Mills field along the loop in the fundamental representation of
the gauge group, $A(\tau)=A_\mu^at^adx_\mu/d\tau$, $\mbox{Tr}(t^at^b)
= \frac{1}{2}\delta^{ab}$. The gauge transformation of $A(\tau)$ is

\beq
A(\tau) \rightarrow S(\tau)A(\tau)S^{-1}(\tau)
+iS(\tau)\frac{d}{d\tau}S^{-1}(\tau).
\la{gt}\eeq
Let $H_i$ be the generators from the Cartan subalgebra ($i=1,...,r;\;
r$ is the rank of the gauge group) and the $r$-dimensional vector ${\bf
m}$ be the eldest weight of the representation in which the Wilson loop
is considered. The formula for the Wilson loop derived in ref.
\cite{DP0} is a path integral over all gauge transformations $S(\tau)$
which should be periodic along the contour:

\beq
L=\int DS(\tau) \exp\left[i\int d\tau\; \mbox{Tr}\; m_iH_i\; (SAS^{-1}
+i S\dot S^{-1})\right].
\la{W1}\eeq
For example, in the simple case of the $SU(2)$ group \eq{W1} reads:

\beq
L=\int DS(\tau) \exp\left[iJ\int d\tau \;\mbox{Tr}\;\tau^3 (SAS^\dagger
+i S\dot S^\dagger)\right]
\la{W12}\eeq
where $\tau^3$ is the third Pauli matrix and $J=\frac{1}{2},\;1,\;
\frac{3}{2},...$ is the `spin' of the representation of the Wilson
loop considered. In what follows we shall concentrate for simplicity
on the $SU(2)$ gauge group.

The path integrals over all gauge rotations \urs{W1}{W12} are not
of the Feynman type: they do not contain terms quadratic in the
derivatives in $\tau$. Therefore, a certain regularization is
understood in these equations: for example, one can introduce
quadratic terms in the angular velocities $iS\dot S^\dagger$ with
small coefficients which eventually should be put to zero; see
ref.\cite{DP0} for details, where \eq{W12} have been derived in two
independent ways. Let us stress that \eqs{W1}{W12} are manifestly gauge
invariant, as is the Wilson loop itself.

The second term in the exponent of \eqs{W1}{W12} is in fact a
Wess--Zumino-type term, and it can be artificially rewritten not as
a line but as a surface integral inside the closed contour of the
Wilson loop. Indeed, let us parametrize the $SU(2)$ matrix $S$ from
\eq{W12} by the Euler angles,

\beq
S=\exp(i\psi \tau^3/2)\;\exp(i\theta\tau^2/2)\;\exp(i\phi\tau^3/2).
\la{Euler}\eeq
The second term in the exponent of \eq{W12} is then

\beq
iJ\;\int d\tau\;\mbox{Tr}(\tau^3iS\dot S^\dagger)
= iJ \;\int d\tau (\cos\theta\;\dot \phi +\dot \psi)
\la{WZ1}\eeq
where the last term is in fact zero due to the periodicity of
$S(\tau)$.

Introducing a unit three vector

\beq
{\n}=\frac{1}{2} \mbox{Tr}\;(S{\bf \tau}S^\dagger \tau^3)
=(\sin\theta\cos\phi, \sin\theta\sin\phi, \cos\theta)
\la{n}\eeq
we can rewrite \ur{WZ1} as

\beq
\ur{WZ1}=i\frac{J}{2}\int d\tau d\sigma \epsilon^{abc}\epsilon_{ij}
n^a\partial_i n^b\partial_j n^c,
\la{WZ2}\eeq
where one integrates over any surface spanned on the contour. Indeed,
the integrand of \eq{WZ2} is known to be a full derivative; using the
Stokes theorem (the standard one!) one reproduces \eq{WZ1}. Let us note
that the r.h.s.  of \eq{WZ2} is the `topological charge' of the ${\n}$
field at the surface:

\beq
Q=\frac{1}{8\pi}\int d\tau d\sigma \epsilon^{abc}\epsilon_{ij}
n^a\partial_i n^b\partial_j n^c.
\la{topc}\eeq

\Eq{WZ1} can be also rewritten in a form which is invariant under
the reparametrizations of the surface. Introducing the invariant
element of a surface,

\beq
d^2\sigma_{\mu\nu}=d\sigma\,d\tau\;\left(
\frac{\partial x_\mu}{\partial \tau}
\frac{\partial x_\nu}{\partial \sigma}-
\frac{\partial x_\nu}{\partial \tau}
\frac{\partial x_\mu}{\partial \sigma}\right)
= \epsilon_{\mu\nu}\;d^2(Area),
\la{elsur}\eeq
one can rewrite \eq{WZ1} as

\beq
\ur{WZ1}=i\frac{J}{2}\int\; d^2\sigma_{\alpha\beta}
\epsilon^{abc} n^a\partial_\alpha n^b\partial_\beta n^c.
\la{WZ3}\eeq

We get for the Wilson loop

\beq
L=\int D{\bf n}(\sigma,\tau)\;\exp\left[iJ\int\,d\tau
(A^an^a)+\frac{iJ}{2}\int\;d^2\sigma_{\alpha\beta}
\epsilon^{abc} n^a\partial_\alpha n^b\partial_\beta n^c\right].
\la{W2}\eeq

The interpretation of this formula is obvious: the unit vector $\n$ plays
the role of the instant direction of the colour "spin" in colour space;
however, multiplying its length by $J$ does not yet guarantee that we
deal with a true quantum state from a representation labelled by $J$ --
that is achieved only by introducing the Wess--Zumino term in \eq{W2}:
it fixes the representation to which the probe quark of the Wilson loop
belongs to be exactly $J$.

Finally, we can rewrite the exponent in this formula so that both terms
appear to be surface integrals:

\beq
L=\int D{\bf n}(\sigma,\tau)\;\exp\frac{iJ}{2}\left(-\int\,
d^2\sigma_{\alpha\beta}F_{\alpha\beta}^an^a+
\int\;d^2\sigma_{\alpha\beta}
\epsilon^{abc} n^a\left(D_\alpha n\right)^b\left(D_\beta
n\right)^c\right),
\la{W21}\eeq
where
$D_\alpha^{ab}=\partial_\alpha\delta^{ab}+\epsilon^{acb}A_\alpha^c$
is the covariant derivative and $F_{\alpha\beta}^a = \partial_\alpha~
A_\mu^a - \partial_\beta~A_\nu^a + \epsilon^{abc}~A_\alpha^b~A_\beta^b$
is the field strength.  Indeed, expanding the exponent of \eq{W21} in
powers of $A_\alpha$ one observes that the quadratic term cancels out
while the linear one is a full derivative reproducing the $A^an^a$ term
in \eq{W2}; the zero-order term is the Wess--Zumino term \ur{topc}.
Note that both terms in \eq{W21} are explicitly gauge invariant. We call
\eq{W21} the new non-abelian Stokes theorem. Another version of a
non-abelian Stokes theorem has been suggested some time ago by Simonov
\cite{Sim}.

One can introduce a gauge-invariant field strength,

\beq
G_{\alpha\beta}=F_{\alpha\beta}^an^a
- \epsilon^{abc} n^a\left(D_\alpha n\right)^b\left(D_\beta
n\right)^c,
\la{gifs}\eeq
which, as a matter of fact, coincides in form with the gauge-invariant
field strength introduced by Polyakov \cite{Pol2} and 't Hooft \cite{tH2}
in connection with monopoles. In that case the unit-vector field $\n$ had
the meaning of the direction of the elementary Higgs field,
$\overrightarrow{\phi}~/ ~|\phi|$.

\section{Quark--monopole interaction}

We start with considering a non-relativistic abelian electric charge $e$
moving in the field of a mag\-ne\-tic monopole sitting at the origin and
having the mag\-ne\-tic field ${\bf B} = (g/4\pi){\r}/|\r|^3$. The equation
of motion for the charge is given by the Lorentz force:

\beq
m\ddot{x_i} = \frac{eg}{4\pi}\epsilon_{ijk}\frac{x_j\dot{x_k}}{|{\bf
x}|^3}.
\la{Lor}\eeq
It is known that the Dirac quantization condition requires $eg=4\pi n$;
we shall choose $n=1$. The relativistic generalization of \eq{Lor},
when the charge is moving along some world line $x_\mu(\tau_1)$ and the
monopole is moving along some world line $y_\mu(\tau_2)$, is

\beq
m\frac{d}{d\tau_1}\frac{\dot x_\mu}{\surd{\dot x_\mu^2}}
= \frac{2}{\pi}\int d\tau_2\epsilon_{\mu\alpha\beta\gamma}
\frac{\left(x(\tau_1)-y(\tau_2)\right)_\alpha}
{\left|x(\tau_1)-y(\tau_2)\right|^4}\frac{dx_\beta(\tau_1)}{d\tau_1}
\frac{dy_\gamma(\tau_2)}{d\tau_2}.
\la{rellor}\eeq
Indeed, taking the monopole sitting at the origin so that its world line
is $y_\mu(\tau_2)=(\tau_2,0,0,0)$, and a non-relativistic charge with
$x_\mu(\tau_1)=(\tau_1,x_i(\tau_1)), \;\;\;\;|\dot x_i|\ll 1$,
we return, after integrating over $\tau_2$ from minus to plus infinity,
to \eq{Lor}. We note that \eq{rellor} is Lorentz invariant and also
invariant under the re-parametrization of both world lines,
$x_\mu(\tau_1)$ and $y_\mu(\tau_2)$, as it should be.

The l.h.s. of \eq{rellor} is obviously the variation of the standard
relativistic action of a free particle,

\beq
S_{free}=m\int d\tau \surd{\dot x_\mu^2};
\la{freeact}\eeq
what about the right-hand side? The appropriate interaction term appears
to be a tricky thing. It can be written only as a non-uniquely defined
Wess--Zumino action. Let us introduce a unit 4-vector

\beq
u_\mu(\tau_1,\tau_2) =
\frac{x_\mu(\tau_1)-y_\mu(\tau_2)}{|x(\tau_1)-y(\tau_2)|},\;\;\;\;
u_\mu^2=1,
\la{u}\eeq
and define its analytical continuation to a third dimension labelled by
$\sigma,\;\;0\leq \sigma \leq 1$, so that

\beq
v_\mu(\tau_1,\tau_2, \sigma=1) = u_\mu(\tau_1,\tau_2),\;\;\;\;
v_\mu(\tau_1,\tau_2, \sigma=0)=  w_\mu,
\la{cont}\eeq
where $w_\mu$ is some constant 4-vector of unit length. One can
also introduce an $SU(2)$ unitary matrix

\beq
V(\tau_1,\tau_2,\sigma)=v_\mu(\tau_1,\tau_2,\sigma)\cdot \sigma_\mu^-,
\;\;\;\; \sigma_\mu^-=(1_2,-i\overrightarrow{\tau})
\la{U}\eeq
where $\overrightarrow{\tau}$ are the three Pauli matrices and
$1_2$ is the unity $2\times 2$ matrix, and define three
anti-hermitean matrices,

\beq
L_A=V^\dagger\partial_A V, \;\;\;\;A=\tau_1,\;\tau_2,\;\sigma.
\la{LA}\eeq

The needed relativistic charge--monopole interaction term can be written
as the winding number of $V$ (times $2\pi$, to make $\exp(iS_{int})$
uniquely defined):

\[
S_{int}=\frac{2\pi}{24\pi^2}\int d\tau_1 d\tau_2 d\sigma \epsilon_{ABC}
\mbox{Tr}\;(L_A L_B L_C)
\]
\beq
= -\frac{1}{6\pi}\int d\tau_1 d\tau_2 d\sigma \epsilon_{ABC}
\epsilon^{\alpha\beta\gamma\delta}\partial_Av_\alpha\partial_Bv_\beta
\partial_Cv_\gamma v_\delta.
\la{WN}\eeq
Indeed, let us vary this $S_{int}$ in respect to the trajectory of the
charge $x_\mu(\tau_1)$. To that end we first find the change of
\ur{WN} due to the arbitrary variation
$\delta v_\mu(\tau_1,\tau_2,\sigma)$. We have

\[
\delta\left[
\epsilon_{ABC}\epsilon^{\alpha\beta\gamma\delta}
\partial_Av_\alpha\partial_Bv_\beta
\partial_Cv_\gamma v_\delta\right]
=
3\epsilon_{ABC}\epsilon^{\alpha\beta\gamma\delta}
\partial_A\delta v_\alpha\partial_Bv_\beta
\partial_Cv_\gamma v_\delta
\]
\beq
+
\epsilon_{ABC}\epsilon^{\alpha\beta\gamma\delta}
\partial_Av_\alpha\partial_Bv_\beta
\partial_Cv_\gamma \delta v_\delta.
\la{var1}\eeq
The last term is zero here since partial derivatives as well as the
variation of $v_\mu$ must be all orthogonal to $v_\mu$ and to each other
(because of the antisymmetric $\epsilon^{\alpha\beta\gamma\delta}$),
which is impossible in four dimensions. Therefore, we are left with the
first term in \eq{var1}, which, for the same reason, can be written as a
full derivative,

\beq
\ur{var1}=3\partial_A\left[\epsilon_{ABC}\epsilon^{\alpha\beta\gamma\delta}
\delta v_\alpha\partial_Bv_\beta
\partial_Cv_\gamma v_\delta\right].
\la{var2}\eeq
Integrals over full derivatives in $\tau_{1,2}$ are zero, if we assume
that the trajectories $x_\mu(\tau_1)$ and $y_\mu(\tau_2)$ are closed, so
that we are left only with the full derivative in the auxiliary dimension
labelled by $\sigma$. Therefore we get

\beq
\delta S_{int}=-\frac{1}{\pi}\int d\tau_1d\tau_2 \int d\sigma
\frac{\partial}{\partial\sigma}\left[\epsilon^{\alpha\beta\gamma\delta}
\delta v_\alpha\partial_{\tau_1}v_\beta
\partial_{\tau_2}v_\gamma v_\delta\right].
\la{var3}\eeq
The value of the square brackets at $\sigma=0$ is zero since we have
chosen the continuation \ur{cont} in such a way that $v_\mu$ at
$\sigma=0$ is equal to a constant vector $w_\mu$
and is thus $\tau_{1,2}$-independent. Therefore, the integral over
$\sigma$ reduces to the value of the square bracket at $\sigma=1$ where
$v_\mu$ assumes its physical value $u_\mu(\tau_1,\tau_2)$, as defined
by \eq{u}.  We have thus demonstrated a well-known general fact that, though
the Wess--Zumino term is not uniquely defined, its variation is.

We now take the variation $\delta v_\mu$ (equal to $\delta u_\mu$ at the
physical surface $\sigma=1$) to be due to the variation of the charge
trajectory $x_\mu(\tau_1)$. In the environment of \eq{var3} it means
that

\beq
\delta v_\mu = \frac{\delta x_\mu(\tau_1)}{|x(\tau_1)-y(\tau_2)|}.
\la{var4}\eeq
We get finally

\beq
\frac{\delta S_{int}}{\delta x_\mu(\tau_1)}
= \frac{2}{\pi}\int d\tau_2\epsilon_{\mu\alpha\beta\gamma}
\frac{\left(x(\tau_1)-y(\tau_2)\right)_\alpha}
{\left|x(\tau_1)-y(\tau_2)\right|^4}\frac{dx_\beta(\tau_1)}{d\tau_1}
\frac{dy_\gamma(\tau_2)}{d\tau_2},
\la{var5}\eeq
that is the r.h.s of the equation of motion \ur{rellor}, {\it q.e.d}.
Actually, \eq{WN} is a generalization of the non-relativistic
charge-monopole interaction, see, e.g., refs.\cite{G,DPbar}.

The main point is that one cannot write down the action whose variation
is given by \eq{var5}, in a unique way. This is because it is impossible
to write the electric charge -- magnetic charge interaction without
introducing a string (or other more complicated objects) which take away
the magnetic flux so that the QED equation, $\mbox{div} {\bf B}=0$, is
satisfied identically. The concrete form of the `string' depends on the
concrete way one parametrizes the continuation of the physical vector
$u_\mu(\tau_1,\tau_2)$ to the unphysical dimension labelled by $\sigma$.
Since the action is given by the `winding number' \eq{WN}, different
parametrizations may differ only by multiples of $2\pi$. Let us stress
that the correct coefficient in \ur{var5} representing the Dirac
quantization condition, follows from the normalization factor
$1/24\pi^2$ of the winding number, see \eq{WN}.

\section{Non-abelian monopoles}

Writing down the charge--monopole interaction in the Wess--Zumino form
we have actually integrated off the gauge fields, leaving only the
charge and monopole trajectories as the dynamical variables of the
theory. Therefore, $\exp(iS_{int})$ is, in fact, what is called the
{\em generating functional} of the theory, depending on the external
charge current $j_\mu^e(x)$ determined by the charge trajectory
$x_\mu(\tau)$. In case of the Yang--Mills theory the external colour
current $j_\mu^e(x)$ is an adjoint matrix. Some time ago we have shown
\cite{DPWZ} that the Yang--Mills generating functional has specific
properties which follow from gauge invariance. It can be written as
a sum of two pieces: one is gauge invariant (call it $W_1$) and can depend
only on the {\em diagonal} part of the current $d_\mu^e$,

\beq
d_\mu^e(\tau)=S(\tau)j_\mu^e(\tau)S^\dagger(\tau)
\la{diagcur}\eeq
(one can always locally rotate the adjoint matrix to make it
diagonal), while the second piece is gauge non-invariant, and depends
on the `angle' variables $S(\tau)$. According to ref.\cite{DPWZ} the
general form of the generating functional in the Yang--Mills theory for
a closed loop of a point-like current can be written as

\beq
W[j_\mu^e]=W_1[d_\mu^e(\tau)]+J\int d\tau S\dot S^\dagger,
\la{genfun}\eeq

In calculating the vacuum average of the Wilson loop one actually has
to substitute the appropriate colour current $j_\mu^e$, as induced by the
specific loop considered, into the generating functional of the theory.
The appropiate current of the Wilson loop can be read off from \ur{W12}
(or \eq{W1} for a general gauge group). It follows from \eq{W12} that the
diagonal part of the probe current is nothing but

\beq
d_\mu^e(x)=2J\int d\tau\;\dot{x}_\mu(\tau)\;
\delta^{(4)}(x_\alpha-x_\alpha(\tau)),
\la{diagcurW}\eeq
where $x_\alpha(\tau)$ is the trajectory of the colour charge, while
its `angle' part $S(\tau)$ is nothing but the gauge transformation
matrix over which one has to integrate in \eq{W12}. Comparing
\eqs{W12}{genfun} one observes that the second
(gauge non-invariant) pieces are {\em exactly cancelled}. This
has been the conclusion of ref. \cite{DPWZ}. Therefore, the integration
over all possible gauge transformations $S(\tau)$ in \eq{W12} becomes
trivial, so that one is left only with the gauge-invariant piece of the
generating functional, $W_1$. If one assumes that the dynamics of large
Wilson loops is governed by monopoles, this $W_1$ is just the
Wess--Zumino-type charge--monopole interaction term, \eq{WN}, where one has
to use the {\em diagonal} part of the colour charge current, given by
\eq{diagcurW}.

It is less evident how to introduce colour monopoles, however one can
think of proceeding in a similar way as with the colour charges. Namely,
the interaction of colour magnetic charges with the {\em dual}
Yang--Mills potential $B_\mu$ is given by a dual Wilson loop (cf.
\eq{W12}),

\beq
L^m=\int DS(\tau)\exp\left[iK\int d\tau \;\mbox{Tr}\;\tau^3 (SBS^\dagger
+i S\dot S^\dagger)\right],
\la{dW}\eeq
where $B(\tau)$ is the tangent component of the dual Yang--Mills
potential along the monopole loop, $S(\tau)$ is the unitary matrix of
the instant colour orientation of the monopole, $K=\frac{1}{2},\;1\;
\frac{3}{2}$,... is the colour `spin' of the monopole. As in the case
of the similar \eq{W12} one has to integrate actually over unit
orientation vectors ${\bf m}(\tau)$ from the coset space, rather than
over the group elements $S(\tau)$.

One can introduce the field theory of monopoles in the path-integral
formalism and write the partition function as a sum over arbitrary
numbers of monopole loop trajectories, with their colour orientation
vectors ${\bf m}(\tau)$ living on those trajectories:

\beq
{\cal Z}^m_K=\sum_n\frac{1}{n!}\prod_i^n
\int D{\bf m}_i(\tau)\int Dy_{i\mu}(\tau)\;
A_K[y_{i\alpha}(\tau)],
\la{PF}\eeq
where $A_K[y_\alpha(\tau)]$ are some weight functions. For the free
particles of mass $m_K$ the weight function is $A_K(l)=\exp(-m_Kl)$ where
$l=\int d\tau\surd\dot{y}_{i\alpha}^2$ is the monopole loop length. With
such a weight functional \eq{PF} is equivalent to the partition
function of the field theory of free scalar Yang--Mills particles with mass
$m_K$ and colour `spin' belonging to a certain representation labelled
by $K$. One can take into account that monopoles can, in principle,
belong to different representations of the gauge group, however for
simplicity we consider only some particular $K$. In a theory of
interacting monopoles the weight functional is, of course, more
complicated.

The main idea of this paper is to avoid the notion of monopole fields
and of dual Yang--Mills potentials (as well as the usual ones), and to
work directly with charge and monopole trajectories. The Wess--Zumino-type
interaction term \ur{WN} solves the problem of eliminating the dual
as well as the usual Yang--Mills potential, therefore it can be
understood as the generating functional -- not only for the electric
charges but also for the magnetic ones, since they enter the
interaction term on equal footing. Therefore, everything said above
about the non-abelian electric currents applies to the non-abelian
magnetic currents. The Wess--Zumino term containing the instant
colour orientation of monopoles (the second term in \eq{dW}) has to be
cancelled exactly by the gauge non-invariant term of the generating
functional similar to that of \eq{genfun}. Then the integration over
${\bf m}_i$ in the monopole partition function \ur{PF} becomes trivial,
and one is left with the {\em abelian} charge--monopole interaction term
where the abelian currents are given by \eq{diagcurW} for the electric
charges and by

\beq
d_\mu^m(x)=2K\int d\tau\;\dot{y}_\mu(\tau)\;
\delta^{(4)}(x_\alpha-y_\alpha(\tau)),
\la{diagcurm}\eeq
for the magnetic currents, where $y_\alpha(\tau)$ are the trajectories
of monopoles.

We arrive thus to a simple recipe to compute the non-abelian electric
Wilson loop in the background of fluctuating quantum monopole fields
-- without introducing such fields at all. It is given by the exponent
of the charge--monopole interaction \ur{WN}, averaged with the monopole
partition function \ur{PF}. We obtain for the averaged Wilson loop whose
trajectory is denoted by $x_\mu(\tau)$:

\[
\langle L[x_\mu(\tau)] \rangle
=\frac{1}{{\cal Z}^m}\sum_n\frac{1}{n!}\prod_i^n
\int Dy_{i\mu}(\tau))\;A_K[y_{i\alpha}(\tau)]
\]
\beq
\times \exp\left\{(i(2J)(2K)S_{int}[x_\mu,y_{i\mu}]\right\}.
\la{NAW}\eeq

The generalization to arbitrary gauge groups and representations of
both charges and mono\-poles is straight\-forward, accord\-ing to
\eq{W1}:  one has to replace $2J\tau^3$ (and $2K\tau^3$) by $\sum
m_iH_i$ where $H_i$ are $r$ generators from the Cartan subalgebra and
${\bf m}$ is the $r$-dimensional eldest weight of the representation to
which the electric charge belongs (and similarly for the magnetic
charge).

\Eq{NAW} relates directly the Wilson loop to the integral over monopole
paths. If confinement, i.e. the area behaviour of the Wilson loop, is
due to the monopole condensation, it should be seen from this formula.
Both monopole condensation and confinement can be thus formulated in
completely gauge-invariant terms: it is a statement about the weight
functional $A_K[y_{i\alpha}(\tau)]$. For example, one can easily check that
the weight $A_K\sim\exp(-ml)$ corresponding to monopoles being free massive
particles, {\em does not} lead to the area law: monopoles should be at least
effectively massless. Long loops are absolutely necessary to derive the
area law from \eq{NAW} \cite{DPloop}.

\section{Conclusions}

We have derived a new non-abelian Stokes theorem for the Wilson loop.
The path ordering of the Wilson loop is exchanged for the integration
over an auxiliary scalar field from the coset $SU(N)/[U(1)]^{(N-1)}$
space; its meaning is the instant direction of the colour charge. With
the help of that field one can construct a gauge-invariant field
strength to which the Stokes theorem applies. The construction needs
an introduction of a non-uniquely defined Wess--Zumino term for the
auxiliary field, which fixes the group representation of the Wilson
loop considered.

We have presented the interaction between quark and monopole
in a Lorentz-invariant, gauge-invariant and reparametrization-invariant
form. It is also given by a Wess--Zumino-type action; its non-uniqueness
is due to the arbitrariness of how one draws the Dirac string. However,
under the change of the string direction it can only produce phase factors
$\exp(2\pi in)=1$.

The average of the quark Wilson loop over the vacuum filled with
fluctuating monopoles has been related directly to path integrals
over monopole trajectories. This enables one to formulate in a gauge
invariant way -- in the language of monopole world lines -- what precisely
does the alleged monopole condensation mean, and how does one get the area
behaviour of the Wilson loop out of that \cite{DPloop}.

\end{document}